\documentclass[journal=jacsat,manuscript=article]{achemso}

\usepackage{indentfirst}
\usepackage[T1]{fontenc} 
\usepackage{gensymb}

\usepackage{achemso}

\author{Richard A. Escalante}
\email{rescalante1@uc.cl}
\author{Mohan C. Mathpal}
\affiliation{Instituto de Física, Ponticia Universidad Católica de Chile, 7820436, Santiago, Chile}
\author{Luis J. Martínez}

\affiliation{Center for Quantum Optics and Quantum Information, Universidad Mayor, Camino La Pirámide 5750, Huechuraba, Chile}

\author{Loïk Gence}
\author{Griselda Garcia}
\affiliation{Instituto de Física, Ponticia Universidad Católica de Chile, 7820436, Santiago, Chile}

\author{Iván A. González}
\affiliation{Departamento de Química, Facultad de Ciencas Naturales, Matemática y del Medio Ambiente, Universidad Tecnológica Metropolitana, Las Palmeras 3360, Ñuñoa, Santiago, Chile}

\author{Jerónimo R. Maze}
\affiliation{Instituto de Física, Ponticia Universidad Católica de Chile, 7820436, Santiago, Chile}
\email{jmaze@puc.cl}

\title[An \textsf{achemso} demo]
  {Optical Characterization of a Single Quantum
Emitter Based on Vanadium Phthalocyanine Molecules}

\begin{document}

\begin{tocentry}
\center
\includegraphics[scale=0.23]{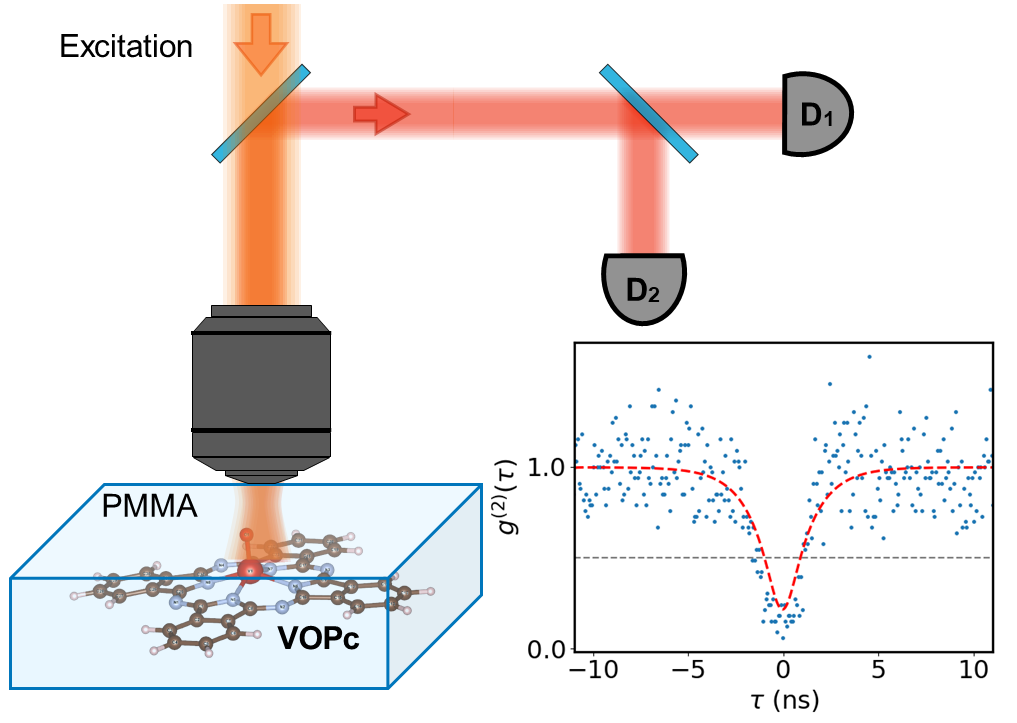}
\end{tocentry}

\begin{abstract}

Single quantum emitters play a fundamental role in the development of quantum technologies such as quantum repeaters, and quantum information processing. Isolating individual molecules with stable optical emission is an essential step for these applications, specially for those molecules that present large coherence times at room
temperature. Among them, vanadium-oxide phthalocyanine (VOPc) molecules stand out as promising candidates due to their large coherence times measured in ensemble. However, the optical properties of individual molecules have not yet been reported. Here we show that single VOPc molecules with stable optical properties at room temperature can be isolated. We find that the optical response of the molecule under laser illumination of different polarization agrees well with a system having pyramidal C$_{4v}$ symmetry. Furthermore, the molecule reveals a non-radiative transition rate that depends on the excitation wavelength when its lifetime is interrogated. We provide theoretical calculations that support our experimental findings and provide insight to the role of phonons and internal electronic structure of the molecule. These results demonstrate that this single paramagnetic molecule can function as a single quantum emitter while displaying optical stability under ambient conditions to have their intrinsic properties investigated.
\end{abstract}
\newpage

The access to stable single emitters is paramount for enabling quantum applications in the optical domain such as quantum information, cryptography and quantum metrology. Several systems have been explored to find such degrees of freedom ranging from quantum dots\cite{KAGAN2021}, crystal defects in large bandgap materials\cite{WOLFOWICZ2021}, and single molecules\cite{TONINELLI2021,BAYLISS2020,FATAFTAH2018}. Among them, the nitrogen-vacancy (NV) center in diamond stands out as a stable single emitter with optical access to an electronic spin with good coherence times even at room temperature\cite{SCHIRHAGL2014}. These defects have allowed exceptional applications in quantum metrology for exploring the electric and magnetic fields emanated from other nanoscale materials\cite{CASOLA2018}. Many of the properties of an NV center rely on the symmetry of its atomic configuration, leading to a unique optical and magnetic dynamics very different from other color centers in diamond\cite{MAZE2011}, and making symmetry a significant feature to consider when choosing a single emitter. However, symmetrical systems are not unique to crystal defects. Several molecules also share equivalent symmetries to those found in crystal defects showing a large variety of zero field splittings and ground and excited state multiplicities\cite{KRZYSTEK2006} and might well be a sufficient characteristic when leading with nanoscale system for quantum applications. Furthermore, some molecules have shown decent coherent times even at room temperature\cite{GRAHAM2017}. Among them, VOPc has a detectable EPR signal\cite{ASSOUR1965}, shows micro second coherence times when probed in ensemble with conventional CW-EPR methods\cite{ATZORI2016} as well as pulsed methods.\cite{BADER2016} Additionally, they have  been shown to couple to microwave photons\cite{BONIZZONI2017}. This motivates the study of this system at the single molecular level. 

Phthalocyanine (Pc) is an organic compound with formula (C$_{8}$H$_{4}$N$_{2}$)$_{4}$, which can be combined with metal cations leading to a wide range of metal-ligand complexes with different electronic and optical properties. Pc has a planar C$_{4h}$ symmetry but when combined with metallic cations such as vanadium-oxide, it acquires a pyramidal planar shape best described with the symmetry point group C$_{4v}$. VOPc molecules show a strong absorption in the 550-950 nm range\cite{HUANG1982} in its solid form and are known for their stability under visible light illumination\cite{GU1995}. Several properties of this molecule have been studied in ensemble form such as nonlinear optical properties\cite{GUO2002} and individual molecules have been imaged by means of low temperature scanning tunneling microscopy\cite{NIU2014} and high-resolution atomic force microscopy\cite{KAISER2019}.

In this paper, we show that single molecules of VOPc can be isolated and their intrinsic properties investigated while embedded in a polymer matrix of PMMA, which has shown to be a reliable medium for single molecule studies.\cite{WILLETS2005, CHEN2012, LEE2012, NOTHAFT2012, TREUSSART2001} These molecules display a very high photo-stability, allowing us to perform several characterization measurements.  We verified the presence of a single VOPc molecule by performing anti-bunching correlation measurements. Saturation measurements resulted in a higher photo-luminescence (PL) intensity under 658 nm excitation compared to 515 nm, consistent with the measured absorption spectrum. Green excitation lifetime measurements resulted in a 0.25 ns lifetime when fitted to a mono-exponential. We also investigated the PL and spectral response to the excitation polarization angle. These measurements fit well with our model which accounts for the orientation of the molecule within the polymer. We compared this to the emission spectrum of a multi-emitter spot, which we observed a more pronounced Gaussian peak centered near 900 nm.

\section{Results and Discussion}

\begin{figure}
  \includegraphics[width=\linewidth]{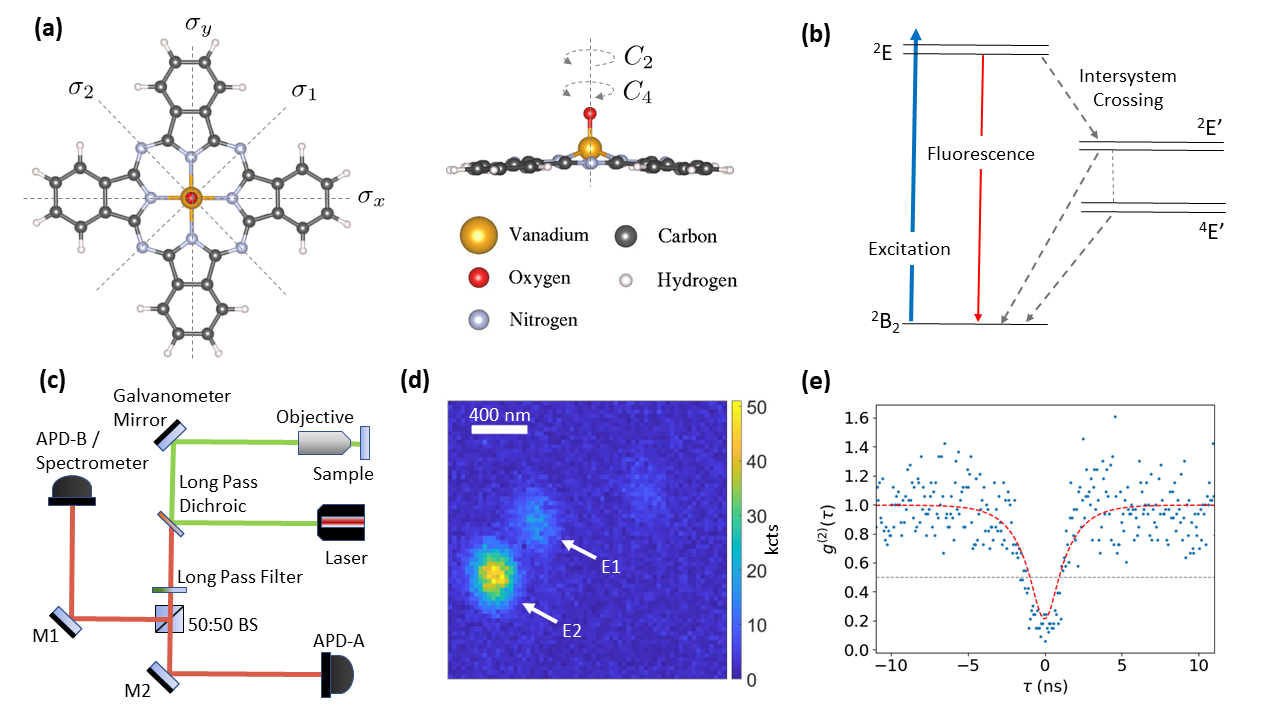}
  \caption{(a) The molecular structure of VOPc showing C$_{4v}$ symmetry. (b) Energy level diagram of VOPc showing some of the decay pathways. (c) Experimental set up of the confocal microscope in a Hanbury-Brown-Twiss configuration. (d) Confocal image showing diffraction limited spots in PMMA. (e) Antibunching measurement fitted with a single exponential verifying the presence of a single emitter (E1). Performing a mono-exponential fit and convoluted with a Gaussian for the IRF, we obtain an excited state lifetime of 1.27 ns. Horizontal grey line indicates the 0.5 threshold for a single emitter.}
  \label{fig:figure1}
\end{figure}

Commercially available powder VOPc and PMMA were purchased (Sigma-Aldrich)and processed as described in the Methods section. This molecule contains a vanadium - oxygen pair at the center of a phthalocyanine molecule. As mentioned before, this lifts the molecular center to create an atomic configuration with C$_{4v}$ symmetry, see Figure~\ref{fig:figure1}a. Figure~\ref{fig:figure1}b shows the energy level diagram for VOPc. The ground state is a doublet having $^{2}B_{2}$ symmetry, and the optically excited state is a doublet with $^{2}E$ symmetry. There are also two possible states which constitute the intersystem crossing: a doublet, $^{2}E^{\prime}$, and a quartet $^{4}E^{\prime}$, which offers a non-radiative decay path back to the ground state.

In order to perform characterization measurements, we used a confocal setup with a 1.25 numerical aperature (NA) oil immersion objective, see Figure~\ref{fig:figure1}c. The optical setup consisted of a Handbury-Brown-Twiss (HBT) configuration for performing antibunching measurements using two Avalanche Photon Detectors (APD), with the option to take optical spectra while imaging. 

Figure~\ref{fig:figure1}d shows a confocal image of two diffraction limited spots which were investigated. First, antibunching measurements were performed on diffraction limited spot E1 under continuous 658 nm wavelength excitation. We obtained a second order correlation function at zero time g$^{2}$(0) < 0.5, confirming the presence of a single emitter, see Figure~\ref{fig:figure1}e. The fit indicated by the red curve consisted of the convolution of the response for a single emitter, a mono-exponential function with a Gaussian normal distribution to take into account the instrument response function (IRF), see Figure S1 of the Supporting Information. The IRF takes into account the timing jitter associated with the system. The data was normalized according to the equation\cite{Beveratos2001}
\begin{equation}
C_{N} =  \frac{c(\tau) }{N_{1}N_{2}\omega T}
\end{equation}
where $c(\tau)$ is the raw coincidence rate, N$_{1,2}$ are the count rates in each detector, $\omega$ is the time bin width, and T is the is the time duration of the measurement. The lifetime associated with the mono-exponential followed from these measurements is of the order of 1.27 ns. 

To further characterize emitter E1, we measured the luminescence as a function of excitation intensity using both 658 nm and 515 nm wavelengths. The laser power was measured before the dichroic and a 20 second time trace was recorded. After subtracting the background, a curve was fitted to the expected response for a two level system \cite{LI2015}:
\begin{equation}
C(P) =  C_{\infty} \frac{P}{P + P_{sat}}
\end{equation}
where $C_{\infty}$ is the saturated photon count rate, $P_{sat}$ is the saturation excitation power, and P is the excitation intensity.

\begin{figure}
  \includegraphics[width=\linewidth]{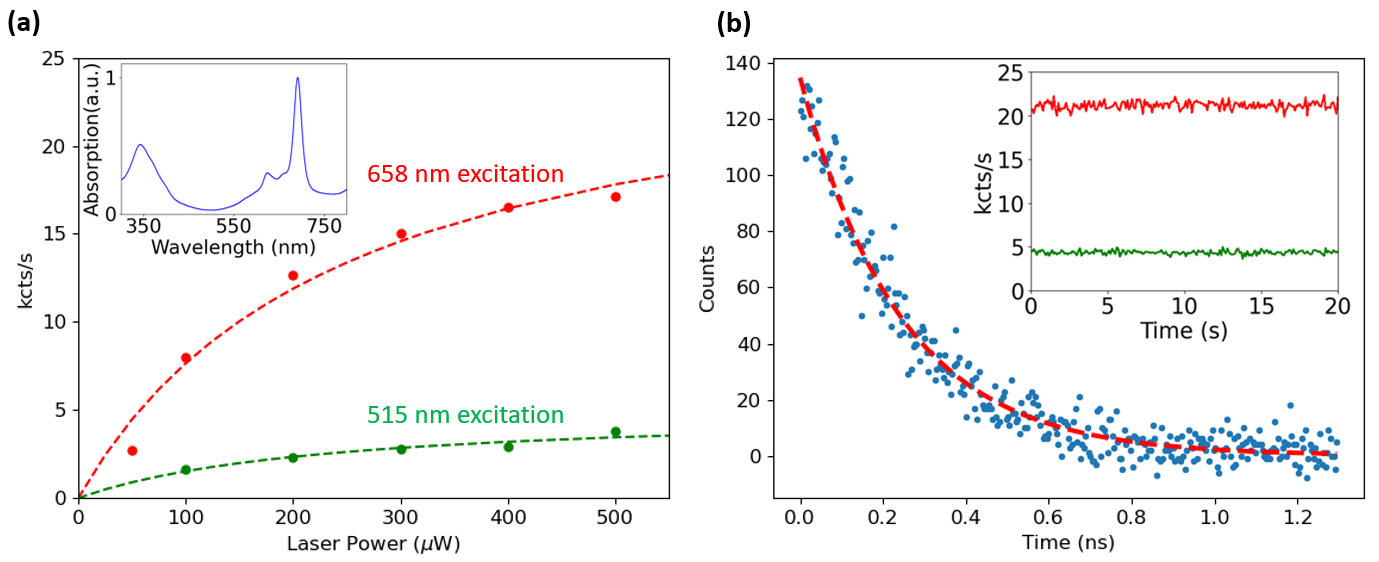}
  \caption{(a) Fluorescence intensity for emitter E1 as a function of laser intensity measured using 515 nm (green) and 658 nm (red) excitation. Fluorescence under 658 nm is seven times larger in agreement with abortion measurements of VOPc dissolved in toluene (inset). We obtained fit values of $C_{\infty} = 26$ kcts/s and $P_{sat} = 249$ $\mu$W for 658 nm, and $C_{\infty} = 5$ kcts/s and $P_{sat} = 232$ $\mu$W for 515 nm. All experiments with 658 nm excitation were done in the $250$ $\mu$W to $300$ $\mu$W range. (b) Lifetime measurements using 515 nm pulsed excitation at 50 Mhz with 50 ps pulse width. Fit shows a mono-exponential decay time of 0.25 ns. Inset: Time trace of emitter E1 under continuous 658 nm (red) and 515 nm (green) excitation at 400 $\mu$W of power, demonstrating high photo-stability under both wavelengths.}
  \label{fig:figure2}
\end{figure}

The observed difference in luminescence under these two excitation wavelengths is consistent with our absorption measurement taken in solution, see the inset in figure~\ref{fig:figure2}a. It can be appreciated that he main absorption bands are centered near 700 nm and 350 nm, and with very little absorption at 500 nm.

Next we performed an additional measurement to interrogate the excited state lifetime using time correlated single photon counting (TCSPC) and a 515 nm pico-second long pulsed laser. Figure~\ref{fig:figure2}b shows the histogram of the time difference between laser pulses and received photons. The associated decay constant ($\tau_{F}$) depends on both the radiative and non-radiative decay channels from the excited state
\begin{equation}
\tau_{F} = (k_{rad} + k_{nr})^{-1}
\end{equation}
where $k_{rad}$ is the radiative rate and $k_{nr}$ is the rate of all the non-radiative processes, including intersystem crossing and internal conversion.\cite{AHARONOVICH2010} Both the second order correlation function g$^{2}(\tau)$ and the decay curve depend on the excited state lifetime. However, in our measurements, the lifetime associated with the correlation function is near 1.27 ns whereas the lifetime measurement using pulsed excitation yields a result of roughly 0.25 ns. This discrepancy can be explained by the increase of non-radiative lifetime under off resonance excitation. The second order correlation function under 658 nm excitation for the g$^{2}(\tau)$ measurement. In this case, the majority of the population decays through radiative channels and the non-radiative rate ($k_{nr}$) is negligible. Meanwhile, for the lifetime measurement using pulsed excitation under 515 nm wavelength excitation, a larger non-radiative rate is expected which results in a shorter measured lifetime ($\tau_{F}$). This non-radiative rate increases for off-resonance excitation has been observed in other systems as well. \cite{BARANOWSKI2019, HATAI1993}

\begin{figure}
  \includegraphics[width=\linewidth]{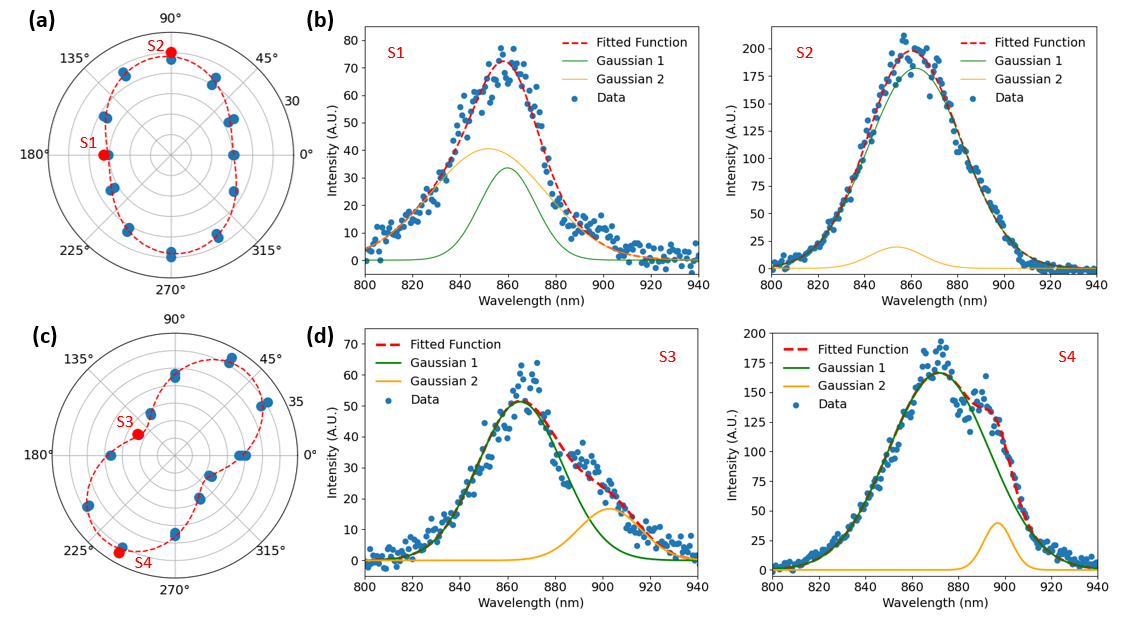}
  \caption{(a) Fluorescence intensity of emitter E1 as a function of the polarization angle. Radial ticks each represent 5 kcts per second with outer most radius being 30 kcts per second. (b) The emission spectrum of E1 taken at an excitation polarization angle of 180\degree (left) and 90\degree (right). (c) Fluorescence intensity of emitter E2 as a function of the polarization angle. (d) The emission spectrum of emitter E2 taken at an excitation polarization angle of 150\degree (left) and 240\degree (right).}
  \label{fig:figure3}
\end{figure}

Next, we observed the emission response as the polarization angle of the excitation beam was rotated. Figure~\ref{fig:figure3}a shows the intensity response of E1. For each intensity data point, we also measured the emission spectrum. Figure~\ref{fig:figure3}b shows the measured spectrum for emitter E1 corresponding to the polarization angles indicated by points S1 and S2 of Figure~\ref{fig:figure3}a . The spectrum fits a two Gaussian centered near 890 nm. We believe this to be the result of vibrational modes and internal structure of the two dipoles associated to the doublet excited state $^{2}E$, which will be discussed later in the text.

We repeated the analysis of rotating the polarization for emitter E2. Figure~\ref{fig:figure3}c shows the emission intensity as a function of the polarization angle. We observe a greater contrast in intensity than emitter E1.In the next paragraphs we explain this effect by considering the inclination of the molecule relative to the excitation beam, i.e., the angle between the normal to the molecular plane and the optical axis, for which the polarization of the light excitation is perpendicular. As it was done with emitter E1, we measured the emission spectrum for each polarization angle. Figure~\ref{fig:figure3}d shows the observed spectra for points S3 and S4. Unlike E1, we see a much more pronounced intensity of the peak centered near 900 nm. The spectrum for each angle can be found in the Supporting Information.

We now discuss the electronic structure of this molecule. As we mentioned before, phthalocyanine alone has a square pyramidal geometry, but when combined with vanadium oxide creates a molecular complex described by the C$_{4v}$ point group. The symmetry operations leaving the molecule invariant are described in the caption of Figure 1a. This point group has four one-dimensional irreducible representations (IRs) named $A_1$, $A_2$, $B_1$, and $B_2$ and one two-dimensional IR named $E$. An \emph{ab-initio} calculation as described in the method section reveals the main symmetrized molecular orbitals (SMO) that participates on the ground and first excited states (see Figure 4a). We have identified the SMOs $a_2$, $b_2$, $e_x$ and $e_y$, named after the irreducible representations $A_2$, $B_2$ and $E$, respectively, they transform to. Others have obtained similar results.\cite{CARLOTTO2018} The ground state configuration as identified by others\cite{KAZMIERCZAK2021} is $^2B_2$ where two paired electrons occupy the $a_2$ state and one electron occupies the $b_2$ state. Interestingly, the $b_2$ state lies lower in energy than the doubly occupied $a_2$ state because the Coulomb interaction of an additional electron on the former state will rise the total energy surpassing that of the latter. Therefore the ground configuration for the ground state is of the form $a_2^2b_2$. The states $e_x$ and $e_y$ are the next lowest unoccupied molecular orbitals (LUMO). The first optically excited state is the result of promoting one electron from the $a_2$ state to either $e_x$ or $e_y$ as illustrated on Figure 4a. This state belongs to the electronic configuration $a_2b_2e$ and therefore we named $^2E$. As the E IR is two dimensional, the excited states is described by two doublets, one with orbital character $e_x$ and another one with orbital character $e_y$, or linear combination of them. The same electronic configuration allows for two other sets of states: two quartets $^4E’$ and another two doublets $^2E’$. These two sets of states constitute the Inter System Crossing (ISC) as a spin flip of one of the electrons is needed to transit from the first optically excited state $^2E$ or to the ground state $^2B_2$ as illustrated by the dashed arrows on Figure 1b. The given names for these states follow from the character of the total electron configuration of the three electrons participating on the different configurations shown in Figure 4a, and following group theoretical considerations. We point out that a mixing between $^2E$ and $^2E'$ will cause transitions between these two sets of excited states and therefore fluorescence from state $^2E'$ as discussed in Abe and Gouterman \cite{AKE1969}.

Upon laser excitation, the observed response as a function of the excitation polarization angle shown on Figure 3a is as expected for a molecule composed of two dipoles. The relevant dipole moments for this optical transition depend on the matrix elements $d_x=\langle e_y|x|a_2\rangle$ and $d_y=\langle e_x|y|a_2\rangle$. Other matrix elements are zero for symmetry reasons. In principle, for a perfect $C_{4v}$ symmetry $d_x=d_y = d$. However, as anticipated before, the polarization response might depend on the relative orientation of the molecular plane with respect to the optical axis. For an angle $\phi$ between the normal vector and the optical axis, the expected fluorescence intensity is given by 
\begin{equation}
I(\theta) \propto d^2\left(1-\sin^2\phi\sin^2\theta\right),
\end{equation}
where $\theta$ is the angle of polarization. For single emitter E1, we find a relative orientation angle $\phi_1 = 37.4^\circ$, meanwhile for the non-single emitter diffraction limited spot E2, $\phi_2 = 53.6^\circ$. We attribute the difference in orientation to the freedom that molecules have to arrange themselves in the PMMA buffer they were prepared during the drop cast of the solution. 

We also investigated the vibronic contributions to the emission spectrum using computer simulations. The results suggest that the molecule is highly symmetric in the ground state and follow C$_{4v}$ symmetry, whereas the change in internal structure partially breaks the symmetry when it goes to the excited state (see Supporting Information Figure S4a and S4b). The excited state geometry along the x-axis partially differs from the geometry along the y-axis as the bond length, bond angles and charges along two axes are slightly different. In agreement with our theoretical observations,  the expected degenerate states split from each other by a small energy gap of about 0.07 eV. Therefore, the fact that the shape of the experimentally observed PL changes as the polarization angle varies can be attributed to the internal geometry and/or symmetry breakage of the molecule in the excited state. As the polarization angle varies, the dipolar strength associated to the mono-electronic transition from the ground state a$_{2}$ to either the excited state e$_{x}$ or e$_{y}$ or a linear combination of both, will also vary, generating slightly different specta as these two excited states will be partially separated in energy. The results indicates that phonon vibrations mainly contribute towards the broadening of the 0-0 transition, which is consistent with the presence of broad intense peak (PL peak at 850 nm). This mode of vibrations in the samples is also experimentally confirmed (Raman peak at 680 cm$^{-1}$) from Raman spectroscopy. For further information on the different vibrations modes that participates on the PL, see the Supporting Information, in particular Figure S3.

\section{Conclusion}

In conclusion, we have isolated a single VOPc molecule, known for its long spin coherence times, in a PMMA polymer and performed optical characterization measurements at room temperature. We observed a very stable PL when excited using 658 nm, and a significant reduction in intensity when using off resonant 515 nm excitation. Although the excited state lifetime measurements differed between the 658 nm continuous wave antibunching and 515 nm pulsed TCSPC measurement, this was likely due to the increased rate of non-radiative processes due to the off-resonance excitation under 515 nm excitation. Finally we measured the PL and emission spectrum as a function of excitation polarization. We compare two diffraction limited spots but with differing PL intensity. The spectrum of the brighter emitter displayed a more pronounced peak at 900 nm, leading us to speculate that this could be a stack of molecules with probably a modified vibrational spectrum and internal energy structure. PL intensity as a function of polarization angle data fits well with our theoretical model which accounts for the angle of inclination the molecule is free to rest within the polymer. We have shown that a quantum emitter can exist in the form of a highly photo-stable single VOPc molecule and have their intrinsic properties investigated while embedded in a polymer matrix. 

\section{Methods}

VOPc powder (commercially available from Sigma-Aldrich) was first dissolved into 2 ml of dimethylformamide (DMF) and further diluted into a solution of dissolved polymethyl methacrylate (PMMA) in toluene. 20 $\mu$l of this mixture was drop casted onto a clean coverslip and heated on a hot plate until dry.

Optical characterization measurements were performed on a home built confocal microscope capable of switching between 515 nm and 658 nm excitation wavelengths as illustrated in Figure 1c.

Anti-bunching measurements to identify single emitters were performed in a HBT configuration under 658 nm continuous wave excitation and approximately 300 $\mu$W of power. Fluorescence was filtered using a dichroic mirror (Semrock FF695-Di02) and a long pass filter (Semrock FF01-715/LP-25). Both APDs (Excelitas SPCM-AQRH-15 and SPCM-AQRH-14-FC) were connected to a time correlation card (Becker and Hickl SPC-130-EM). An oil immersion objective (Nikon Plan Achromat 100x Oil) was used to focus the excitation beam onto the sample and read the luminescence.

For lifetime measurements, a pico-second 515 nm laser (Becker and Hickl BDS-515-SM) was used to apply short pulses at 50 MHz. This laser was also used in continuous wave mode for green saturation curve measurements.

Excitation polarization measurements were done by placing a half wave plate (Thorlabs WPHSM05-670) before the objective. Rotation of the incident polarization was performed using a half wave plate on a motorized rotation mount. For each measurement, the wave plate was rotated $15\degree$ and a 20-second time trace was recorded and an emission spectrum was recorded with 30 seconds integration time (using a spectrometer QE Pro from Ocean Optics). 

Absorption measurements of a solution of VOPc were taken using a spectrophotometer (Shimadzu UV-1900) at 5\degree C. This solution consisted of VOPc dissolved in DMF (concentration of 500 $\mu$g ml$^{-1}$), of which 100 $\mu$l was further diluted into 2.5 ml of toluene.

To corroborate our experimental findings an \emph{ab-initio} calculation based on density functional theory (DFT) methods and time dependent-DFT (TD-DFT) for single molecule systems were performed by using Gaussian 16 package (Revision C.01).\cite{G16} The geometry optimization at the ground state was performed by choosing B3LYP and PBEPBE beased DFT methods and using 6-311G (+, 2d, p) basis set, where a diffuse function and polarization function on heavy atoms was used along with a polarization function on hydrogen atoms. 

\section{Acknowledgements}

Authors acknowledge support from ONR Project N62909-18-1-2180. R.Escalante and J. Maze acknowledges support from ANID-FONDECYT 1121512 and AFOSR FA2386-21-1-4125. L. Gence and J. Maze acknowledges support from ANID-Anillo ACT192023. J. Maze and M. Mathpal greatly acknowledge the funding from FONDECYT 3190316 to support the presented computational research activity. In this work M. Mathpal also acknowledges the support from Prof. Claudio Lopez of the department of chemistry in our university to perform Raman spectroscopy measurements. The authors acknowledge support from ANID-FONDEQUIP projects EQM140168 and EQM180180. L. Martinez acknowledges support from FONDECYT Regular grant No. 1190447. I. Gonzalez acknowledges support from FONDECYT 11180185. 

\section{Supporting Information Available}

Additional experimental information and raw spectrum data, simulation results related to symmetry, Raman data.

\begin{figure}
  \includegraphics[width=\linewidth]{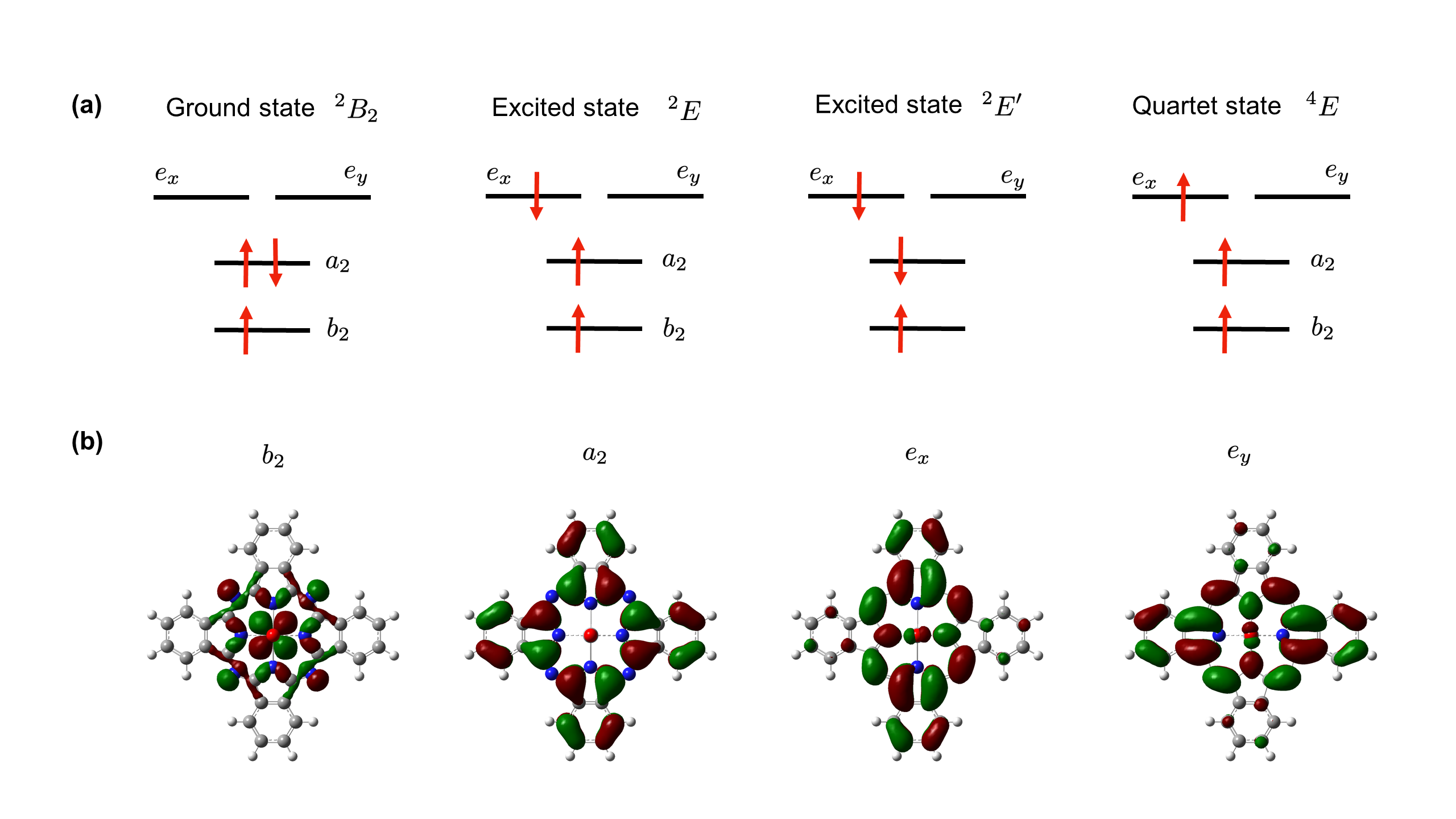}
  \caption{(a) Schematic of the many electron configuration states in a VOPc molecule. (b) Isosurfaces of the symmetrized single-electron molecular orbitals calculated by DFT using the 6-311G and B3LYP method. The name for each orbital is given according to the irreducible representation in transforms to.}
  \label{fig:figure4}
\end{figure}

\newpage
\bibliographystyle{plain} 
\bibliography{UCDraft_bib} 

\end{document}


\section{Instrument Response Function}
To properly fit the $g^{2}(\tau)$ correlation function, we measured the instrument response function (IRF) of the system illustrated in Figure 1 of the main text, including the two avalanche photon detectors (APDs) and the time correlated single photon counting card. To record the IRF, the optical filter was removed and the sample was replaced by a blank coverslip. Next, pico-second long pulses of 515 nm wavelength were applied and the response of the APDs was recorded with the time correlation card, following the procedure in Martínez \emph{et al.}\cite{MARTINEZ2016} Figure S1 shows on of our measured response functions. The red curve is a Gaussian fit, where its width, $\sigma$, represents the time uncertainty. For our system, $\sigma = 0.41$ ns.

Next, in order to obtain the lifetime of our single emitter, the fitted IRF was convoluted with the expected response for a single emitter, $g^2(\tau) = 1 - ae^{-\frac{t}{\tau_{1}}}$, so our $g^{2}_{fit}(\tau)$ is

\begin{equation}
g^{2}_{fit}(\tau) = \int_{-\infty}^{\infty}g^{2}(\tau)G(\tau - t)\, dt
\end{equation}
where $g^2(\tau) = 1 - ae^{-\frac{t}{\tau_{1}}}$ is the antibunching signal of an ideal quantum emitter and $G(\tau - t) = \frac{1}{\sigma\sqrt{2\pi}}e^{\frac{-(\tau - t)^2}{2\sigma^2}}$ is the IRF Gaussian due to the timing jitter of the system. The values we used for the normalization of the $g^2(\tau)$ are $N_1 = 15320$ cts/s, $N_2 = 15360$ cts/s, $T = 1875.99 s$ and $ \omega = 7.33 \times 10^{-11} s$

\begin{figure}
  \includegraphics[scale=0.8]{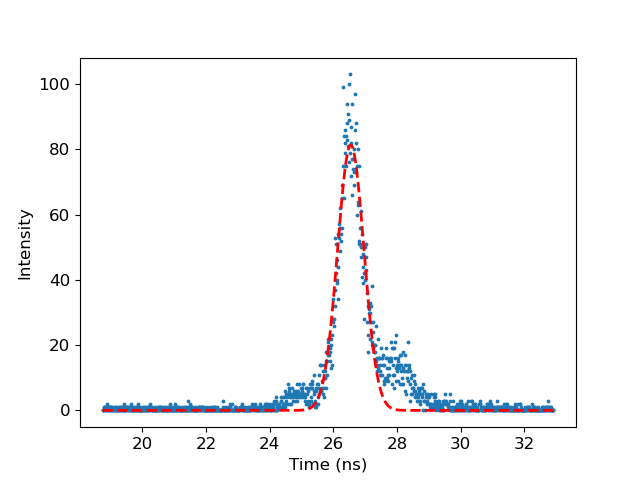}
  \caption{Instrument response function measured using 515 nm picosecond laser pulses. The red curve shows the Gaussian fit used to convolve the g$^{2}(\tau)$ signal. The relevant fitting parameter is $\sigma = 0.41$ ns for the standard deviation. }
  \label{fig:S1}
\end{figure}
\section{Spectrum Data}

We present our recorded spectra data for emitter E1 and E2 as measured for each of the polarization's angle of the excitation beam in Figures S2 and S3, respectively. 

\begin{figure}[H]
  \includegraphics[width=\linewidth]{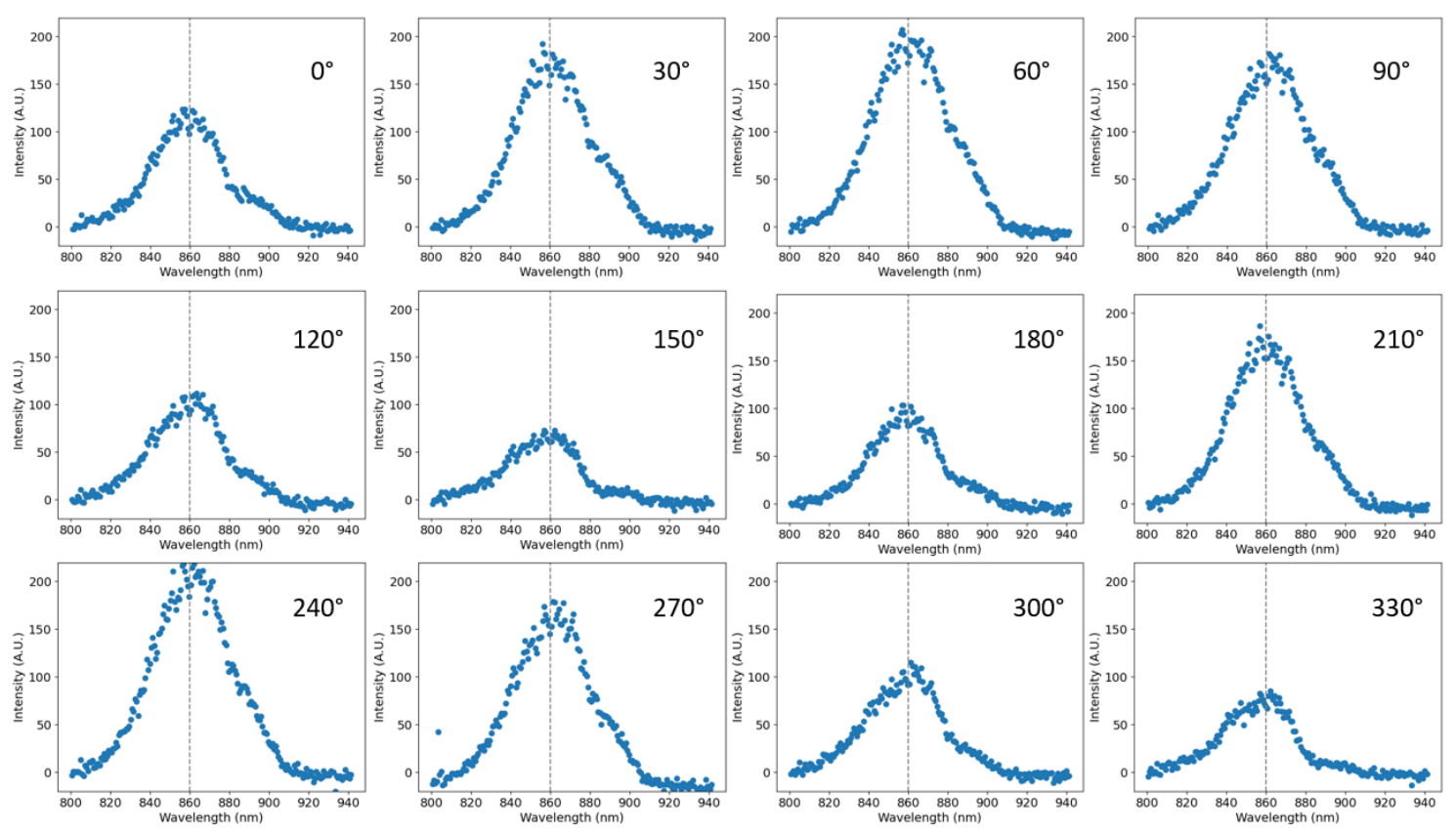}
  \caption{The measured spectrum data for each polarization angle for emitter E1. The dashed grey line marks 860 nm.}
  \label{fig:S2}
\end{figure}
\begin{figure}[H]
  \includegraphics[width=\linewidth]{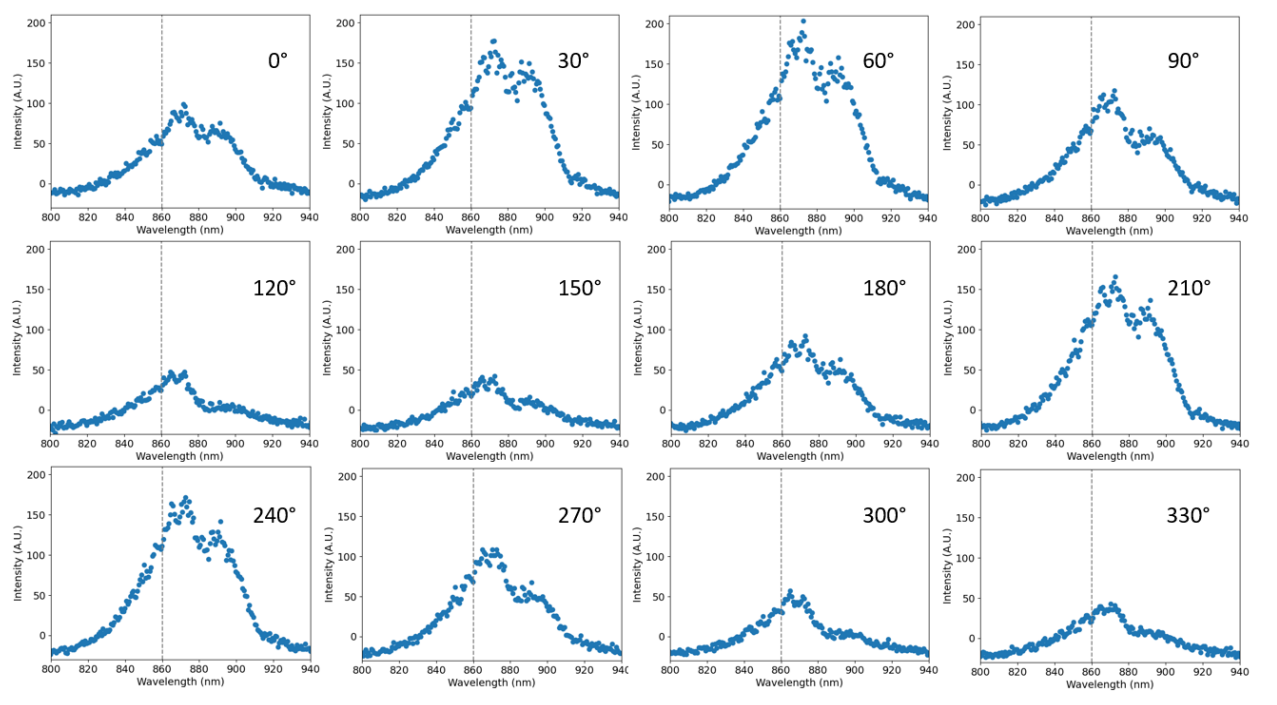}
  \caption{The measured spectrum data for each polarization angle for emitter E2.}
  \label{fig:S3}
\end{figure}

\section{Symmetry Breakage}

In the following, we provide details on the results that support the symmetry breakage of the molecule. Figure S4 shows the bond lengths and charge localization for the ground and excited states. The Mulliken charges are displayed on the top of mainly central atoms, whereas the bond length (in Å) is displayed across (or perpendicular) the length of different bonds. The bond angles (in degrees) are represented with an angle symbol. The molecule is highly symmetric in the ground state and follows C$_{4v}$ symmetry, which is in good agreement with the results reported in the literature.\cite{DEBNATH2022} The excited state geometry was optimized the excited states with the highest oscillator strength. The change in internal structure of the molecule in its excited state illustrates that the VOPc molecule is a good example of a partial symmetry breakage upon excitation (see Figure S4). The optimization of other excited states lead to similar changes in the internal structure. Such changes in molecular geometry can have significant impacts on the electronic and optical properties of the molecules. 

\begin{figure}[H]
  \includegraphics[scale=0.5]{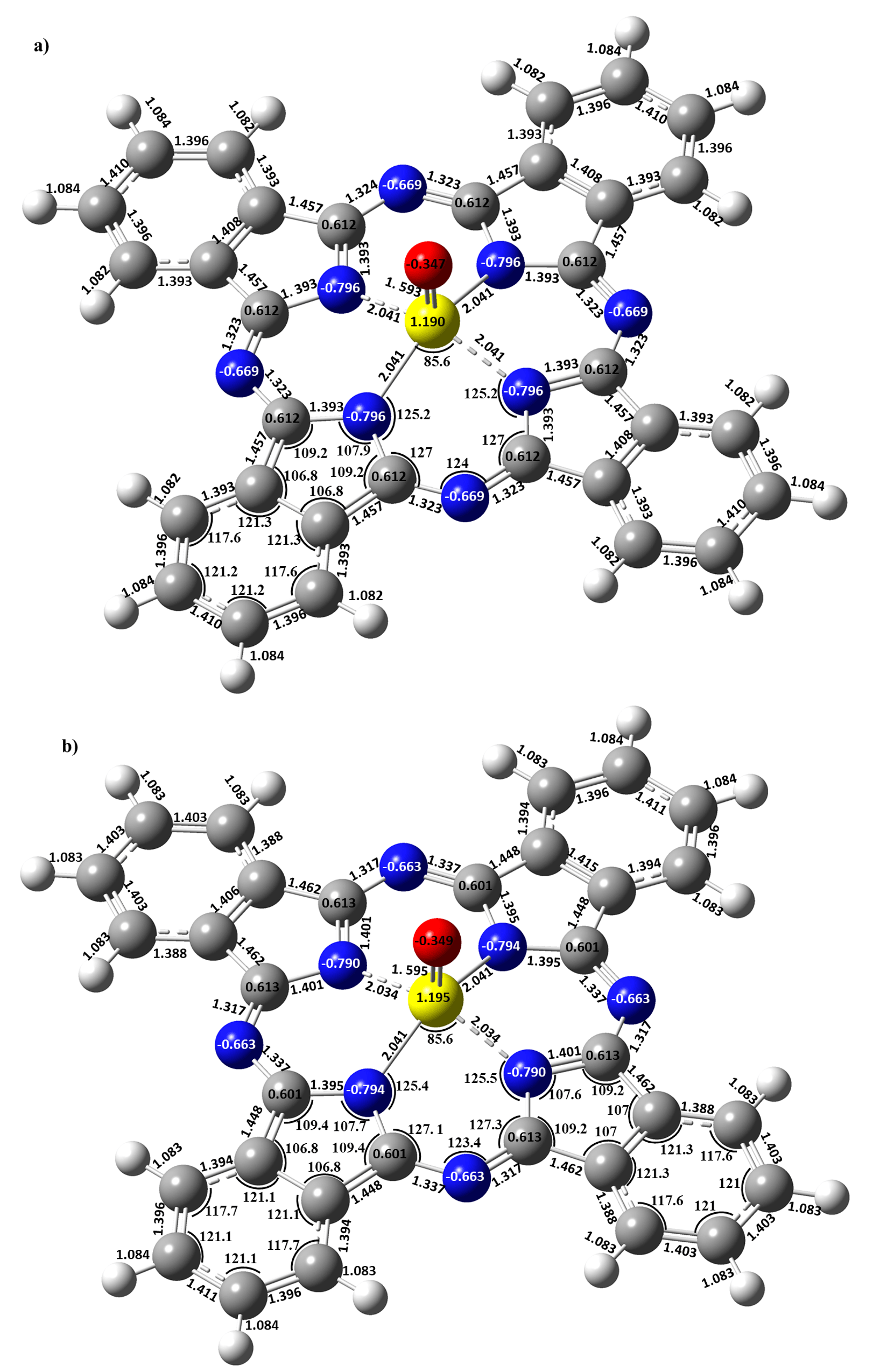}
  \caption{Geometry of VOPc molecule for doublets (a) ground state and (b) excited states calculated by using 321 G basis set and B3LYP DFT method.}
  \label{fig:S4}
\end{figure}

\section{Vibrational Modes}

The theoretical vibronic emissions calculations with Frank-Condon-Herzberg-Teller (FCHT) method (321 G basis set) suggests that there are mainly 3 phonon vibrations that could couple to the photoluminescence emissions, whose frequencies are located at 42 cm$^{-1}$, 692 cm$^{-1}$, 1602 cm$^{-1}$ respectively. The experimentally observed PL emission peak centered about 861 nm is consistent with the zero-phonon line broadened by coupling with phonon vibrations of frequencies 42 cm$^{-1}$ as discussed by N.P. Kazmierczak \emph{et al.}\cite{KAZMIERCZAK2021} A vibration around 1602 cm$^{-1}$ is believed to extend the tail of PL emission in the near-IR region, whose experimentally observed intensity is negligible and to the best of our literature survey this mode has not been reported so far in a VOPc molecule.\cite{DEBNATH2022, KAZMIERCZAK2021} The different vibration modes that couple to the PL emissions in the ground states are displayed in figure S5 (a, b, c). These two modes of vibrations in the samples are also experimentally confirmed from Raman spectroscopy (Raman shift at 680 cm$^{-1}$ and 1526 cm$^{-1}$ respectively. See figure S6.

\begin{figure}[H]
  \includegraphics[scale=1.0]{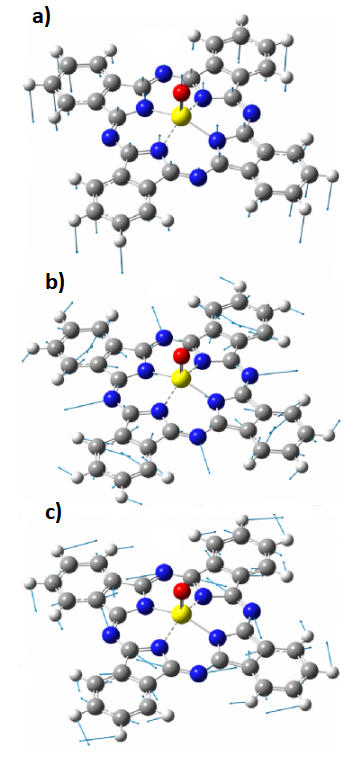}
  \caption{Vibrational motion of the phonon coupling in the ground state (displacement vector is magnified by 10 times for visualization). (a) 42 cm$^{-1}$ (b) 680 cm$^{-1}$ and (c) 1600 $^{-1}$.}
  \label{fig:S5}
\end{figure}

\section{Raman Spectrum}

Figure S6 shows the Raman spectra recorded by using a laser with an excitation wavelength of 532 nm and 10x objective with a focused laser spot size of 4-5 $\mu$m. The peaks that correspond to VOPc molecules in the main sample (image (d)) are assigned to their different possible mode of vibration with the help of theoretical calculations.
\begin{figure}[H]
  \includegraphics[scale=1.0]{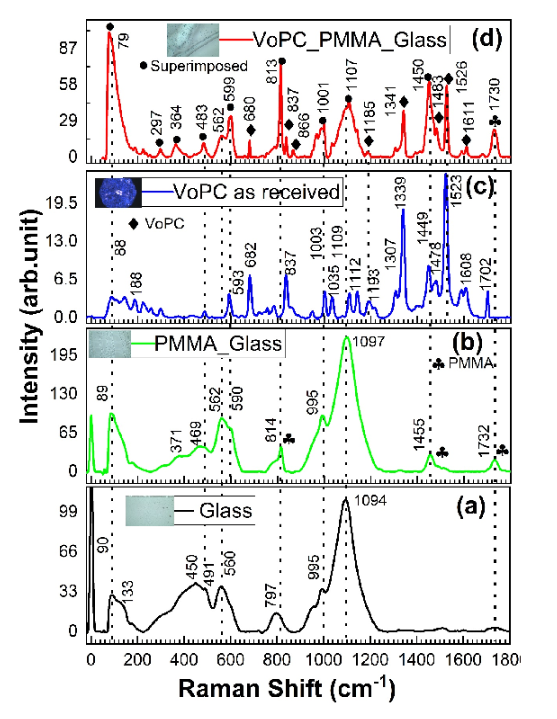}
  \caption{Raman spectra of the VOPc molecules performed under identical conditions. (a) Only glass substrate for reference, (b) PMMA coated on glass substrate for reference. (c) VOPc powder sample as received from Sigma Aldrich company for comparison, and (d) VOPc molecules in PMMA drop cast over glass converslip.}
  \label{fig:S6}
\end{figure}

\bibliographystyle{plainnat}
\bibliography{UCDraft_bib} 